\documentclass[prl,letter,twocolumn,showpacs,lengthcheck]{revtex4}
\usepackage{graphicx}
\usepackage{amsmath,amssymb}
\usepackage{bm}
\usepackage{ifthen}

\newcommand{\tr}{\operatorname{Tr}}

\newcommand{\ket}[1]{| #1 \rangle}

\newcommand{\dr}{{\,\rm d}{\bm r}}
\renewcommand{\vec}[1]{{\bm #1}}

\begin{document}
\title{Electric field control and optical signature of entanglement in
quantum dot molecules}
\author{Gabriel Bester}
\author{Alex Zunger}
\affiliation{National Renewable Energy Laboratory, Golden CO 80401}
\date{\today}

\begin{abstract}
The degree of entanglement of an electron with a hole in a vertically coupled 
self-assembled dot molecule is shown to be tunable by an external 
electric field. Using 
atomistic pseudopotential calculations 
followed by a  configuration interaction many-body treatment of correlations, 
we calculate the electronic states, degree of entanglement and optical
absorption. We offer a novel way to 
spectroscopically detect the magnitude of electric field needed to maximize
the entanglement. 
\end{abstract}

\pacs{
78.67.Hc, 
73.21.La, 
03.67.Mn 
}

\maketitle

A pair of quantum dots or ``quantum dot molecule'' (QDM) occupied by
two electrons \cite{divincenzo95,loss98} or by an electron-hole pair
\cite{bayer01} have been offered \cite{divincenzo95,loss98,bayer01}
as a basis for
quantum computing. The fundamental requirement for such
quantum algorithm is the availability of entangled states and the ability to
entangle and disentangle the quantum bits (qubits). In the context of a dot
molecule, an entangled
electron-hole pair can be represented by the the maximally entangled Bell
state  
$e_Th_T$ + $e_Bh_B$, where $e$ and $h$ stand for the electron and the hole
(the two qubits) and $T$ and $B$ for their localizations in top or bottom
dot. The original proposal \cite{bayer01} and subsequent experiments 
\cite{bayer01,hinzer01,bayer02b} for entangled electron-hole pairs  in QDMs 
promised a {\em high} degree of entanglement \cite{bayer01} based on analysis 
via simple models.
However, later theoretical work showed \cite{bester04a} that electron-hole  
entanglement is generally low in such cases and develops a sharp maximum 
only at a 
specific interdot separation that critically depends on the size difference
of the two dots. Unfortunately, it has proven to be difficult to 
experimentally control so
precisely the interdot distance and the size difference of the two dots.
The question we address here is whether the degree of
entanglement can be maximized by other means, more accessible
experimentally than a variation of the interdot separation.
We propose and quantify theoretically that it is possible to {\em tune} and 
control the degree of entanglement 
by applying an external electric field in the growth direction
\cite{fry00,shtrichman02b,sugisaki02,alen03}. The use of electric field has
been demonstrated in
quantum dots \cite{fry00,shtrichman02b,sugisaki02,alen03} and very recently in
a single quantum dot molecules by 
Krenner {\em et al.} \cite{krenner05}. We predict that, while 
the entanglement at zero field is
generally low (35\% for our case), it can reach a high value 
(75\% in our case) at a specific electric field $F_{Smax}$ 
(-5.4 kV/cm in our case). Moreover, precisely at this field the first two 
exciton lines merge, giving a well defined spectroscopic signature of the 
point of maximum entanglement.
 
In order to obtain reliable results for the correlated exciton states, it is
of foremost importance to accurately account for the
multi-band character of the hole states and for the correct strain dependence
in the coupling region (between the dots). 
We have thus solved the pertinent Schr\"odinger equation atomistically, in a
multi-band fashion. We use  the Hamiltonian 
$H = -1/2\nabla^2 + \sum_{\alpha,n} v_\alpha(\vec{r}-\vec{R}_n) + V_{SO} 
+ |e|Fz $ under an external electric field $F$ applied in [001]
($z$) direction. The atomistic pseudopotentials $v_\alpha$ of atom of type
$\alpha$ and the  
non-local spin-orbit potential $V_{SO}$ are fit to 
reproduce InAs and GaAs bulk properties \cite{williamson00,bester04a}. 
The atomic positions $\{\vec{R}_n\}$ are obtained by minimizing the atomistic 
strain energy (via valence force field \cite{keating66}) for a given shape and size of the 
dots. Our quantum dots have a truncated cone shape (12 nm base and 2 nm
height) with a composition ranging from pure InAs at the top
to In$_{0.5}$Ga$_{0.5}$As at the base, as determined in Ref. \cite{bayer01}.
The single-particle Hamiltonian is diagonalized in a basis 
$\Psi = \sum_{n,k}A_{n,k}\phi_{n,k}$
of pseudopotential Bloch functions $\phi_{n,k}$ as outlined in
Ref. \onlinecite{wang99b}, thus permitting coupling of various Bloch states. 
Correlations are treated via a many-body expansion in Slater
determinants \cite{Szabo89,franceschetti99} where the electrons not included
dynamically are represented by a model screening of the Coulomb and (long and
short range) exchange \cite{resta77}. The entanglement is calculated according
to the von Neumann entropy of entanglement \cite{bennett96,bester05b}. 

\begin{figure}
\centerline{
\includegraphics[width=.48\columnwidth]{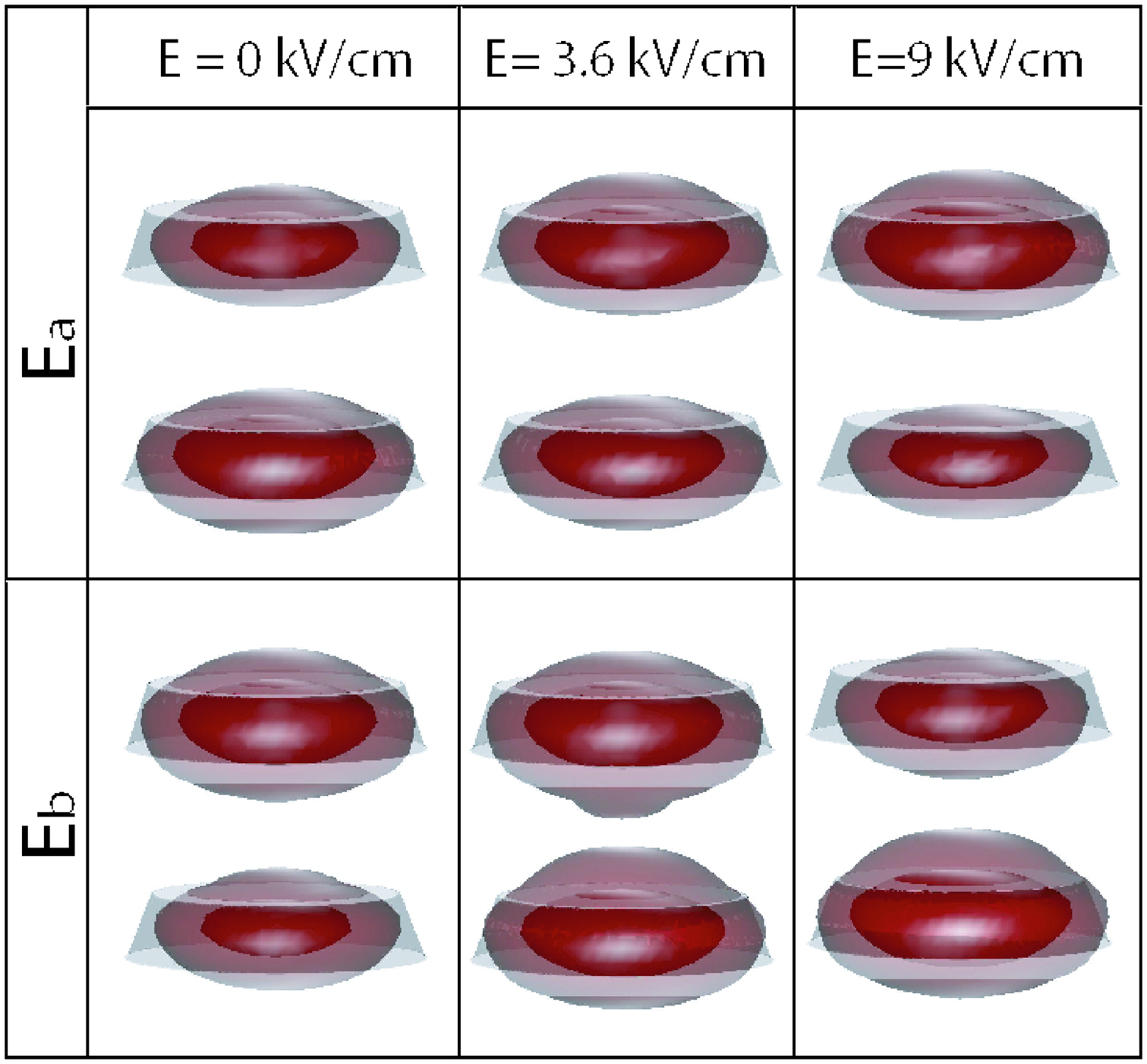}
\includegraphics[width=.48\columnwidth]{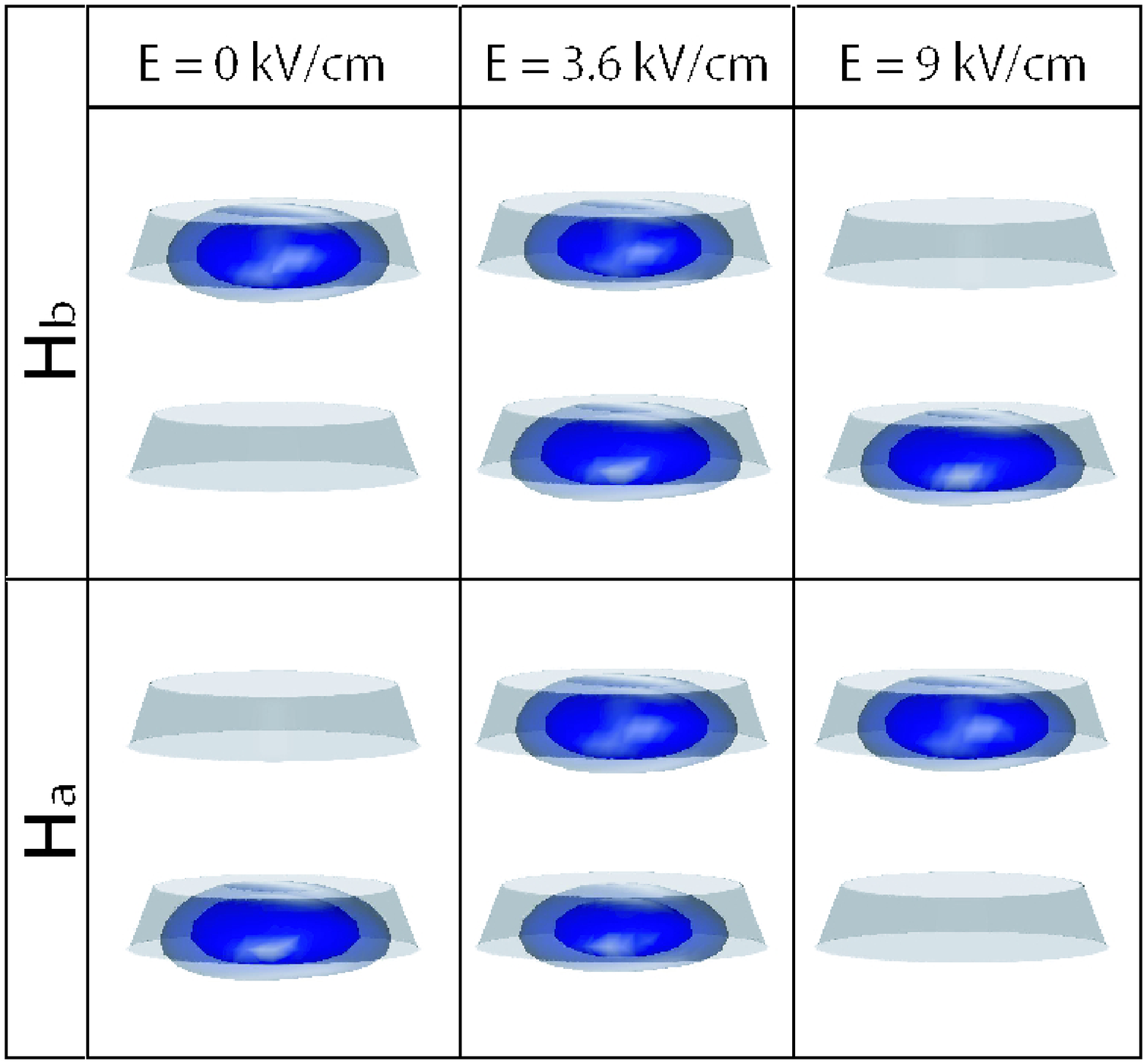}}
\caption{Square of the first two electron and first two hole wave functions of
an InGaAs/GaAs quantum dot as a function of electric field at an interdot 
separation $d$=7.9 nm.}
\label{fig:wfn}
\end{figure}

\begin{figure}
\centerline{\includegraphics[width=.8\columnwidth,angle=0]{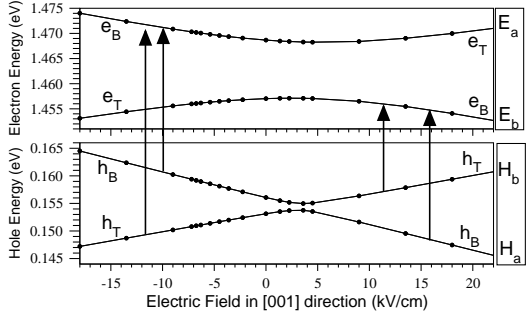}}
\caption{Single particle electron (upper panel) and hole (lower panel) 
eigenvalues as a function of electric field for $d$=7.9 nm with respect to the
GaAs valence band maximum. We indicate the localization of the MOs on top and bottom dot 
by $e_T$, $e_B$, $h_T$, $h_B$.}
\label{fig:sp_energies}
\end{figure}

The bonding ($b$) and antibonding ($a$) electron molecular levels of a dot
molecule will be denoted as $E_{a}$, $E_{b}$. For an idealized (mostly
unrealistic) symmetric case 
the lowest energy molecular orbitals (MOs) develop from  
single-dot electron states $e_{T}$ and $e_{B}$ located  on the  
bottom ($B$) and top ($T$) dots:
\begin{equation}
\label{eq:BA-picture}
\psi [E_{b}] = \frac{1}{\sqrt{2}} (e_{T} + e_{B}); \quad
\psi [E_{a}] = \frac{1}{\sqrt{2}} (e_{T} - e_{B})  ,
\end{equation}
and similarly for the holes $H_{a}$, $H_{b}$.
As shown previously \cite{he05b,bester04a}, in reality, because of strain and random-alloy 
fluctuations, one does
not have a {\em symmetric} bonding-antibonding behavior even if the dot
molecule is made of identical (but non-spherical) dots. This is seen in
Fig.~\ref{fig:wfn} where both electron and hole molecular orbital wave
functions are shown for zero electric field $\vec{F}=0$. 
We see that the (lighter-mass) electrons tunnel between dots, forming
bonding-antibonding states as in Eq.  (\ref{eq:BA-picture}), but the
(heavier-mass) holes remain localized on the top (bottom) dot
for the bonding (antibonding) MO $H_{b}$ ($H_{a}$).
The single-particle molecular orbital energy levels are shown in Fig.~\ref{fig:sp_energies}.
As we apply an electric field the molecular levels that develop from the
single-dot orbitals exhibit anti-crossing. We have indicated in Fig.~\ref{fig:sp_energies}
the major character of the molecular states $E_{a}$, $E_{b}$, $H_{a}$, 
$H_{b}$ in terms of the localization on individual dots ($e_T$, $e_B$,
$h_T$ and $h_B$) using the calculated
MO wave functions of Fig.~\ref{fig:wfn}. We see that for holes at positive fields 
$\psi(H_{a}) \simeq h_B$ and $\psi(H_{b}) \simeq h_A$ while for electrons
$\psi(E_{a}) \simeq e_T$ and $\psi(E_{b}) \simeq e_B$. The opposite is true
for negative fields. Thus, by applying an electric field we can tune the
localization of the MO's and, for instance, compensate for size, composition 
or shape differences of both dots. We will see that 
this tuning of localization will also control the degree of entanglement.

\begin{figure}
\centerline{\includegraphics[width=.95\columnwidth]{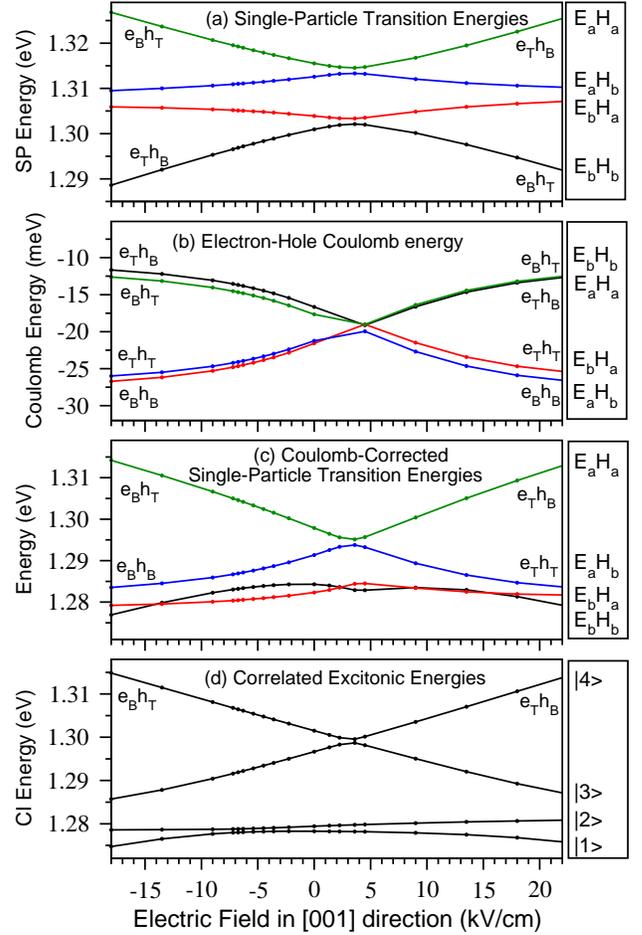}}
\caption{(Color) Transition energies for $d$=7.9 nm as a function of
electric field in different approximations: (a) Single particle transition
energies $\varepsilon_g$ (b) Direct electron-hole Coulomb matrix elements
between MOs $J_{e,h}$. 
(c) Transition energies including electron-hole Coulomb interaction: 
$\varepsilon_g^{i,j} + J_{e,h}^{i,j}$, but without correlation effects. 
(d) Final correlated exciton results. The bidot 
products ($e_Th_T$, etc...) are given whenever the MO states 
(given on the right) are strongly dominated by one of these
products.}
\label{fig:FromMOtoCI}
\end{figure}

There are four transitions between the four molecular levels shown as vertical
arrows in Fig.~\ref{fig:sp_energies}. 
Their single-particle transition
energies $\varepsilon_g^{i,j}$ (differences between the 
energies from Fig.~\ref{fig:sp_energies}) are given
in Fig.~\ref{fig:FromMOtoCI}(a) and show maxima and minima vs field. We note in Fig.~\ref{fig:FromMOtoCI}(a) 
the character of the four transitions in terms of localization on single-dot 
orbitals. We see that at high fields, the lowest- and highest-energy transitions involve 
{\em different} dots: for example $E_{b}H_{b}$ is $e_Bh_T$ at positive fields, 
and $e_Th_B$ at negative fields. Thus, the corresponding dipole transitions are 
expected to be weak (``dark states''). In contrast  the second and third
transitions at high fields involve the {\em same} dots: for example $E_{b}H_{a}$ is 
$e_Bh_B$ at positive fields and $e_Th_T$ at negative field. Thus, the
corresponding  dipole transitions are expected to be large (``bright
states'').
 
The single-particle approximation underlying Fig.~\ref{fig:FromMOtoCI}(a) is valid 
only in the case of large fields, where $e$-$h$ Coulomb effects are small
compared to the field-driven variation in the single-particle levels. At these
large fields, the excitons are pure determinants with localization on 
either $e_Th_T$  or $e_Th_B$ or $e_Bh_T$ or $e_Bh_B$
and therefore show no entanglement. 
We will next see that in the interesting region of electric fields, $e-h$ Coulomb
interactions are crucial. In
Figure \ref{fig:FromMOtoCI}(b) we show the calculated electron-hole Coulomb 
interaction 
\begin{equation}
\label{eq:Js}
J_{e,h}[i-j] =
\int\!\!\!\int \frac{\psi_i^\star(\vec{r}_a)\psi_j^\star(\vec{r}_b)
                \psi_{j}(\vec{r}_b)\psi_{i}(\vec{r}_a)}{
\epsilon(\vec{r}_a,\vec{r}_b)|\vec{r}_a-\vec{r}_b|} {\dr}_a{\dr}_b
\end{equation}
between the MOs ($i=E_{a}$ or $E_{b}$) and ($j=H_{a}$ or $H_{b}$) using
the model of Resta \cite{resta77} for the screening $\epsilon$.
This  interaction $J_{e,h}$ (Fig. \ref{fig:FromMOtoCI}(b)) 
shows the reverse behavior vs field
compared with the MO energies $\varepsilon_g$ vs field
[Fig.~\ref{fig:FromMOtoCI}(a)]. For example, whereas $J_{e,h}[E_{b}H_{b}]$ is
{\em maximal} (less negative) at large positive or negative fields, and {\em minimal}
at intermediate fields, the MO band gap 
$\varepsilon_g [E_{b}H_{b}]$ is {\em minimal} at large positive or 
negative fields and {\em maximal} at intermediate fields. 
Not surprisingly, when one calculates the
Coulomb-corrected excitonic transition energy $\varepsilon_g^{i,j} + J_{e,h}^{i,j}$
between the molecular states $i$ and $j$ [Fig.~\ref{fig:FromMOtoCI}(c)] one sees a partial
cancellation for the two low-energy transitions, $E_{b}H_{b}$ and 
$E_{b}H_{a}$, leading to a weak dependence of the transition energy on
field.  In contrast, inspection of Figs.~\ref{fig:FromMOtoCI}(a) and 
\ref{fig:FromMOtoCI}(b) for the two
highest-energy transitions shows that the field-dependence of 
$\varepsilon_g^{i,j}$ and  $J_{e,h}^{i,j}$ {\em reinforce} each other, so the
excitonic gap $\varepsilon_g^{i,j} + J_{e,h}^{i,j}$ [Fig.~\ref{fig:FromMOtoCI}(c)] has an {\em
amplified} dependence on field. We conclude that the combination of $\varepsilon_g$
and $J_{e,h}$ brings the lowest-energy transitions {\em closer} to each other,
while pushing the two higher-energy transitions {\em apart}. This will affect
the correlation coupling between the MO's, as seen next. 

The Coulomb-corrected excitonic transition energies 
$\varepsilon_g^{i,j} + J_{e,h}^{i,j}$ neglect the  interactions
between the different configurations, i.e., the states 
from Fig.~\ref{fig:FromMOtoCI}(c) are not
allow to interact. This interaction is included in the next step via a 
configuration interaction (CI)\cite{Szabo89,franceschetti99} calculation in 
which we include all Coulomb and exchange integrals from the first four electron
and first four hole states (including spin). The results are shown in 
Fig.~\ref{fig:FromMOtoCI}(d)  as a function of 
electric field. The lowest energy transitions (excitons $\ket{1}$ and 
$\ket{2}$) have a very weak dependence on field, similarly to the case
without correlations [Fig.~\ref{fig:FromMOtoCI}(c)]. In contrast to 
the perturbative approach of Fig.~\ref{fig:FromMOtoCI}(c) the states do not
cross but anticrossing at -5.4 kV/cm, as expected from  interacting states. 
Similarly, $\ket{3}$ and $\ket{4}$ anticross at +3.6 kV/cm but in a more abrupt
fashion.

\begin{figure}
\centerline{\includegraphics[width=.95\columnwidth]{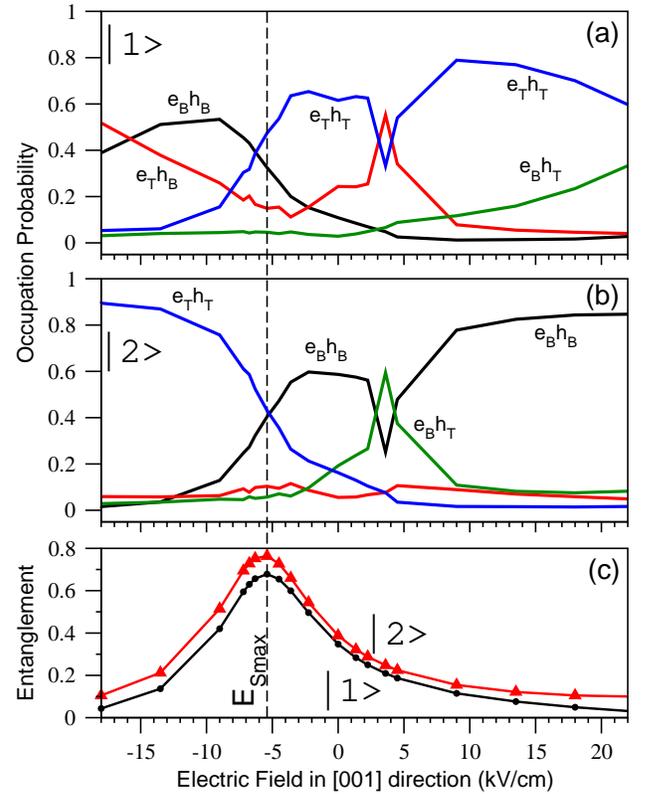}}
\caption{Occupation probabilities given as the bidot products $e_Th_T$ or
$e_Th_B$ or $e_Bh_T$ and $e_Bh_B$ as a function of the electric field
for the first two exciton states $\ket{1}$ (a) and $\ket{2}$ (b) for $d$=7.9
nm. (c) Entropy of entanglement of the first two excitons.}
\label{fig:occupations}
\end{figure}

\begin{figure}
\centerline{\includegraphics[width=.97\columnwidth]{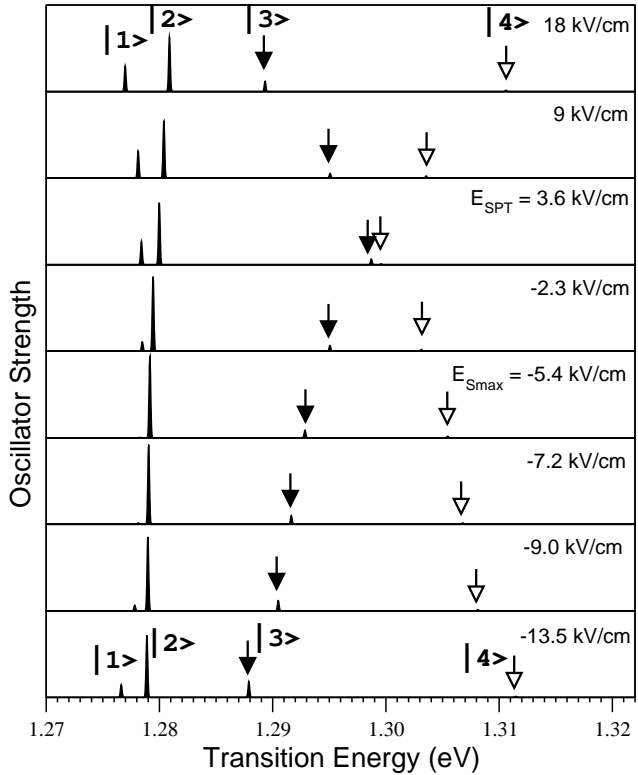}}
\caption{Oscillator strength of the first four transitions (first 16
transitions including spin) as a function of electric field for $d$=7.9 nm.}
\label{fig:spectrum} 
\end{figure}

To understand the correlated CI results we next analyze  $\ket{1}$, $\ket{2}$, $\ket{3}$,
$\ket{4}$ by decomposing the correlated excitonic states into sums of products
of single-dot states $e_Th_T$, $e_Bh_T$, $e_Th_B$ and  $e_Bh_B$ called ``bidot
products'' (see Ref. \onlinecite{bester05b} for details). The results of 
this decomposition are given in 
Fig.~\ref{fig:occupations}(a)(b) for states $\ket{1}$ and $\ket{2}$ as 
a function of the electric field. 
Fig.~\ref{fig:occupations}(c) shows the degree of entanglement
calculated for the correlated CI wave functions using the von Neumann
formula \cite{bennett96,bester04a} 
$S=-\tr \rho_{\text{e}} \log_2 \rho_{\text{e}}$, where $\rho_{\text{e}}$ is
the reduced density matrices of the electron. When a state is made solely of a
single bidot product such as $e_Bh_T$ it is unentangled, but when it is made
of a coherent superposition, such as $e_Bh_B \pm e_Th_T$, it might be 
entangled. In the case of very strong positive fields 
(larger than 20 kV/cm), state $\ket{1}$ is purely $e_Bh_T$ (unentangled), 
as the field pulls the electron to the
bottom dot and the hole to the top dot. The excitons $\ket{2}$ and 
$\ket{3}$ are already, at moderate positive fields, purely $e_Bh_B$ and 
$e_Th_T$, respectively, and remain unentangled states at large fields. 
The entanglement of $\ket{1}$ and $\ket{2}$ [Fig.~\ref{fig:occupations}(c)] 
reaches its maximum of around 75\% at $F_{Smax}$ = -5.4 kV/cm.  At this field  
the  exciton states $\ket{1}$ and $\ket{2}$ are mainly composed of 
$e_Th_T \pm e_Bh_B$ configurations, as shown in Fig.~\ref{fig:occupations}(a)(b)
\footnote{In order for this e-h entanglement to be
useful in a quantum computation scheme, the qubits need to be physically
separated to be addressed  individually. One possibility for such a separation
seems to be the  use of {\em in plane} electric fields that would coherently
drive the electron and hole to different neighboring quantum dots. This would
constitute the next experimental challenge}. The calculated excitonic states of
Fig.~\ref{fig:FromMOtoCI}(d) are now used to calculate the absorption spectra
in Fig.~\ref{fig:spectrum}. In Fig.~\ref{fig:spectrum} the oscillator strength
is plotted for different values of the electric field as a function of the transition
energy. The plot shows a total of four transitions marked with $\ket{1}$, $\ket{2}$, 
$\ket{3}$, $\ket{4}$. Since the transitions $\ket{3}$ and $\ket{4}$ are weak,
their position is marked by solid and open arrows respectively. 
The transitions $\ket{1}$ and $\ket{2}$ ($\ket{3}$ and $\ket{4}$) have a
weak (strong) dependence on field and show an anticrossings at $E_{\rm Smax}$ 
($E_{\rm SPT}$). The spectra show that at $E_{\rm Smax}$ the lowest energy
exciton $\ket{1}$ becomes dark and progressively gains oscillator strength 
away from the anti-crossing. The point of merging of $\ket{1}$ and
$\ket{2}$ at the  field $E_{Smax}$ reflects a  ``resonant conditions'' with 
maximum entanglement (Fig.~\ref{fig:occupations}). The distinct spectroscopic
signal of the anticrossing (where the lowest line progressively looses
oscillator strength) occurs at the point of maximum entanglement and, we
suggest, can
give experimentalists a simple way to control a delicate quantity such as 
entanglement. 

\begin{figure}
\centerline{\includegraphics[width=\columnwidth]{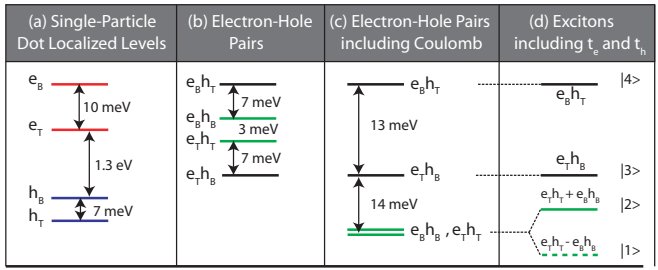}}
\caption{Formation of the highly entangled excitonic states at the critical
field F=-5.4 kV/cm. (a) Single particle electron and hole levels in the
dot-localized basis. (b) Simple differences between the single particle
electron and hole energies from (a). (c) Adding electron-hole direct Coulomb
interaction to (b). (d) Adding electron and hole hopping the the levels
from (c)}
\label{fig:model} 
\end{figure}

The physics underlying the  ``resonant condition'', that produces the high
degree of entanglement, is revealed in Fig.~\ref{fig:model} using the
more intuitive basis of dot-localized orbitals (as opposed to MO's). 
To obtain dot-localized orbitals, we rotate the
MO's (that are delocalized over both dots) until the on-site Coulomb 
interaction is maximized \cite{edmiston63}. This procedure yields
single-particle energies of the {\em dot-localized} orbitals, denoted
as $e_T$, $e_B$, $h_T$ and $h_B$ in Fig.~\ref{fig:model}(a). 
Fig.~\ref{fig:model}(a) shows that at $F_{Smax}$ the energies of the
dot-localized electron and hole orbitals are separated by 10 and 7 meV
respectively with a gap of 1.3 eV. Generally, the energy separation between $e_T$ and 
$e_B$ and between $h_T$ and $h_B$ can reflect size, composition or shape 
differences of the two dots. These differences can be tuned by the electric
field. Fig.~\ref{fig:model}(b) shows the energies
of simple products of these electron and hole states. They show two closely spaced
levels (3 meV apart) in the center of the spectrum and two states 
7 meV lower and higher in energy. These energies are different from
the MO's energies of Fig.~\ref{fig:FromMOtoCI}(a) that are combinations of 
$e_Th_T$, $e_Th_B$, $e_Bh_T$ and $e_Bh_B$ at this field. In the next step, in
Fig.~\ref{fig:model}(c), the two-body
Coulomb attraction is taken into account and lowers the $e_Th_T$ and $e_Bh_B$ 
states in such a way that they are about 14 meV below the $e_Th_B$
state. This is the consequence of a weak $e-h$ binding for the dissociated 
excitons $e_Th_B$ and $e_Bh_T$. Notably, the simple products $e_Th_T$ and
$e_Bh_B$ are energetically nearly {\em degenerate} at this level, this is the
``resonant condition'' mentioned above.
In the last step, in Fig.~\ref{fig:model}(d), the excitons $\ket{1}$ and  $\ket{2}$ 
are now created by including the effects of electron and hole hopping that
effectively produce correlated states. The excitons $\ket{1}$ and 
$\ket{2}$ are now a bonding- and antibonding-like combination of the 
energetically degenerate $e_Th_T$ and $e_Bh_B$ states. The excitons 
$\ket{1}$ and  $\ket{2}$ are
now split by a small energy of less than 1 meV. This small splitting is 
conceptually very similar to the Davydov splitting \cite{Craig68} in 
molecular crystals. The analysis also reveals that  
$\ket{1}$ is anti-symmetric ($e_Th_T$ - $e_Bh_B$) and therefore optically 
dark while $\ket{2}$ is symmetric ($e_Th_T$ + $e_Bh_B$)  and optically bright.
The high symmetry of these states (purely symmetric and anti-symmetric)
leads to the high degree of entanglement. Any deviations from $F_{Smax}$ will
lead to a less symmetric combinations  as
$\frac{1}{\sqrt{2}} (\alpha e_Th_T + \beta e_Bh_B)$ with $\alpha \neq \beta$
with lower entanglement and smaller oscillator strength.

In conclusion, we  showed that the degree of electron-hole 
entanglement in coupled  quantum dots can be tuned by 
an external electric field and that the point of maximum entanglement can be
identified by measuring  the photoluminescence spectra, observing the merging
of two peaks. This opens new ways for experimentalists to identify the
electric field needed to achieve maximum entanglement in specific dot
molecules. We finally analyzed the nature of the excitons and revealed the 
interplay of single particle effects, direct Coulomb binding and electron 
and hole hopping on the many body levels. We described how these effects
conspire to yield a highly entangled state.  

\acknowledgments
This work was supported by the US Department of Energy (SC/BES) under 
the LAB 03-17 initiative.


\begin{thebibliography}{21}
\expandafter\ifx\csname natexlab\endcsname\relax\def\natexlab#1{#1}\fi
\expandafter\ifx\csname bibnamefont\endcsname\relax
  \def\bibnamefont#1{#1}\fi
\expandafter\ifx\csname bibfnamefont\endcsname\relax
  \def\bibfnamefont#1{#1}\fi
\expandafter\ifx\csname citenamefont\endcsname\relax
  \def\citenamefont#1{#1}\fi
\expandafter\ifx\csname url\endcsname\relax
  \def\url#1{\texttt{#1}}\fi
\expandafter\ifx\csname urlprefix\endcsname\relax\def\urlprefix{URL }\fi
\providecommand{\bibinfo}[2]{#2}
\providecommand{\eprint}[2][]{\url{#2}}

\bibitem[{\citenamefont{DiVincenzo}(1995)}]{divincenzo95}
\bibinfo{author}{\bibfnamefont{D.~P.} \bibnamefont{DiVincenzo}},
  \bibinfo{journal}{Science} \textbf{\bibinfo{volume}{270}},
  \bibinfo{pages}{255} (\bibinfo{year}{1995}).

\bibitem[{\citenamefont{Loss and DiVincenzo}(1998)}]{loss98}
\bibinfo{author}{\bibfnamefont{D.}~\bibnamefont{Loss}} \bibnamefont{and}
  \bibinfo{author}{\bibfnamefont{D.~P.} \bibnamefont{DiVincenzo}},
  \bibinfo{journal}{Phys. Rev. A} \textbf{\bibinfo{volume}{57}},
  \bibinfo{pages}{120} (\bibinfo{year}{1998}).

\bibitem[{\citenamefont{Bayer et~al.}(2001)\citenamefont{Bayer, Hawrylak,
  Hinzer, Fafard, Korkusinski, Wasilewski, Stern, and Forchel1}}]{bayer01}
\bibinfo{author}{\bibfnamefont{M.}~\bibnamefont{Bayer}},
  \bibinfo{author}{\bibfnamefont{P.}~\bibnamefont{Hawrylak}},
  \bibinfo{author}{\bibfnamefont{K.}~\bibnamefont{Hinzer}},
  \bibinfo{author}{\bibfnamefont{S.}~\bibnamefont{Fafard}},
  \bibinfo{author}{\bibfnamefont{M.}~\bibnamefont{Korkusinski}},
  \bibinfo{author}{\bibfnamefont{Z.~R.} \bibnamefont{Wasilewski}},
  \bibinfo{author}{\bibfnamefont{O.}~\bibnamefont{Stern}}, \bibnamefont{and}
  \bibinfo{author}{\bibfnamefont{A.}~\bibnamefont{Forchel1}},
  \bibinfo{journal}{Science} \textbf{\bibinfo{volume}{291}},
  \bibinfo{pages}{451} (\bibinfo{year}{2001}).

\bibitem[{\citenamefont{Hinzer et~al.}(2001)\citenamefont{Hinzer, Bayer,
  McCaffrey, Hawrylak, Korkusinski, Stern, Wasilewski, Fafard, and
  Forchel}}]{hinzer01}
\bibinfo{author}{\bibfnamefont{K.}~\bibnamefont{Hinzer}},
  \bibinfo{author}{\bibfnamefont{M.}~\bibnamefont{Bayer}},
  \bibinfo{author}{\bibfnamefont{J.~P.} \bibnamefont{McCaffrey}},
  \bibinfo{author}{\bibfnamefont{P.}~\bibnamefont{Hawrylak}},
  \bibinfo{author}{\bibfnamefont{M.}~\bibnamefont{Korkusinski}},
  \bibinfo{author}{\bibfnamefont{O.}~\bibnamefont{Stern}},
  \bibinfo{author}{\bibfnamefont{Z.~R.} \bibnamefont{Wasilewski}},
  \bibinfo{author}{\bibfnamefont{S.}~\bibnamefont{Fafard}}, \bibnamefont{and}
  \bibinfo{author}{\bibfnamefont{A.}~\bibnamefont{Forchel}},
  \bibinfo{journal}{Phys. Stat. Sol. (b)} \textbf{\bibinfo{volume}{224}},
  \bibinfo{pages}{385} (\bibinfo{year}{2001}).

\bibitem[{\citenamefont{Bayer et~al.}(2002)\citenamefont{Bayer, Ortner,
  Larionov, Timofeev, Forchel, Hawrylak, Hinzer, Korkusinski, Fafard, and
  Wasilewski}}]{bayer02b}
\bibinfo{author}{\bibfnamefont{M.}~\bibnamefont{Bayer}},
  \bibinfo{author}{\bibfnamefont{G.}~\bibnamefont{Ortner}},
  \bibinfo{author}{\bibfnamefont{A.}~\bibnamefont{Larionov}},
  \bibinfo{author}{\bibfnamefont{V.}~\bibnamefont{Timofeev}},
  \bibinfo{author}{\bibfnamefont{A.}~\bibnamefont{Forchel}},
  \bibinfo{author}{\bibfnamefont{P.}~\bibnamefont{Hawrylak}},
  \bibinfo{author}{\bibfnamefont{K.}~\bibnamefont{Hinzer}},
  \bibinfo{author}{\bibfnamefont{M.}~\bibnamefont{Korkusinski}},
  \bibinfo{author}{\bibfnamefont{S.}~\bibnamefont{Fafard}}, \bibnamefont{and}
  \bibinfo{author}{\bibfnamefont{Z.}~\bibnamefont{Wasilewski}},
  \bibinfo{journal}{Physica E} \textbf{\bibinfo{volume}{12}},
  \bibinfo{pages}{900} (\bibinfo{year}{2002}).

\bibitem[{\citenamefont{Bester et~al.}(2004)\citenamefont{Bester, Shumway, and
  Zunger}}]{bester04a}
\bibinfo{author}{\bibfnamefont{G.}~\bibnamefont{Bester}},
  \bibinfo{author}{\bibfnamefont{J.}~\bibnamefont{Shumway}}, \bibnamefont{and}
  \bibinfo{author}{\bibfnamefont{A.}~\bibnamefont{Zunger}},
  \bibinfo{journal}{Phys. Rev. Lett.} \textbf{\bibinfo{volume}{93}},
  \bibinfo{pages}{47401} (\bibinfo{year}{2004}).

\bibitem[{\citenamefont{Fry et~al.}(2000)\citenamefont{Fry, Itskevich, Mowbray,
  Skolnick, Finley, Barker, O'Reilly, Wilson, Larkin, Maksym et~al.}}]{fry00}
\bibinfo{author}{\bibfnamefont{P.~W.} \bibnamefont{Fry}},
  \bibinfo{author}{\bibfnamefont{I.~E.} \bibnamefont{Itskevich}},
  \bibinfo{author}{\bibfnamefont{D.~J.} \bibnamefont{Mowbray}},
  \bibinfo{author}{\bibfnamefont{M.~S.} \bibnamefont{Skolnick}},
  \bibinfo{author}{\bibfnamefont{J.~J.} \bibnamefont{Finley}},
  \bibinfo{author}{\bibfnamefont{J.~A.} \bibnamefont{Barker}},
  \bibinfo{author}{\bibfnamefont{E.~P.} \bibnamefont{O'Reilly}},
  \bibinfo{author}{\bibfnamefont{L.~R.} \bibnamefont{Wilson}},
  \bibinfo{author}{\bibfnamefont{I.~A.} \bibnamefont{Larkin}},
  \bibinfo{author}{\bibfnamefont{P.~A.} \bibnamefont{Maksym}},
  \bibnamefont{et~al.}, \bibinfo{journal}{Phys.\ Rev. \ Lett.}
  \textbf{\bibinfo{volume}{84}}, \bibinfo{pages}{733} (\bibinfo{year}{2000}).

\bibitem[{\citenamefont{Shtrichman et~al.}(2002)\citenamefont{Shtrichman,
  Metzner, Gerardot, Schoenfeld, and Petroff}}]{shtrichman02b}
\bibinfo{author}{\bibfnamefont{I.}~\bibnamefont{Shtrichman}},
  \bibinfo{author}{\bibfnamefont{C.}~\bibnamefont{Metzner}},
  \bibinfo{author}{\bibfnamefont{B.~D.} \bibnamefont{Gerardot}},
  \bibinfo{author}{\bibfnamefont{W.~Y.} \bibnamefont{Schoenfeld}},
  \bibnamefont{and} \bibinfo{author}{\bibfnamefont{P.~M.}
  \bibnamefont{Petroff}}, \bibinfo{journal}{Phys. Rev. B}
  \textbf{\bibinfo{volume}{65}} (\bibinfo{year}{2002}).

\bibitem[{\citenamefont{Sugisaki et~al.}(2002)\citenamefont{Sugisaki, Ren,
  Nair, Nishi, and Masumoto}}]{sugisaki02}
\bibinfo{author}{\bibfnamefont{M.}~\bibnamefont{Sugisaki}},
  \bibinfo{author}{\bibfnamefont{H.~W.} \bibnamefont{Ren}},
  \bibinfo{author}{\bibfnamefont{S.~V.} \bibnamefont{Nair}},
  \bibinfo{author}{\bibfnamefont{K.}~\bibnamefont{Nishi}}, \bibnamefont{and}
  \bibinfo{author}{\bibfnamefont{Y.}~\bibnamefont{Masumoto}},
  \bibinfo{journal}{Physical Review B} \textbf{\bibinfo{volume}{66}}
  (\bibinfo{year}{2002}).

\bibitem[{\citenamefont{Alen et~al.}(2003)\citenamefont{Alen, Bickel, Karrai,
  Warburton, and Petroff}}]{alen03}
\bibinfo{author}{\bibfnamefont{B.}~\bibnamefont{Alen}},
  \bibinfo{author}{\bibfnamefont{F.}~\bibnamefont{Bickel}},
  \bibinfo{author}{\bibfnamefont{K.}~\bibnamefont{Karrai}},
  \bibinfo{author}{\bibfnamefont{R.}~\bibnamefont{Warburton}},
  \bibnamefont{and} \bibinfo{author}{\bibfnamefont{P.}~\bibnamefont{Petroff}},
  \bibinfo{journal}{Appl. Phys. Lett.} \textbf{\bibinfo{volume}{83}},
  \bibinfo{pages}{2235} (\bibinfo{year}{2003}).

\bibitem[{\citenamefont{Krenner et~al.}(2005)\citenamefont{Krenner, Sabathil,
Clark, Kress, Schuh, Bichler, Abstreiter, Finley}}]{krenner05}
\bibinfo{author}{\bibfnamefont{H.J.}~\bibnamefont{Krenner}},
  \bibinfo{author}{\bibfnamefont{M.}~\bibnamefont{Sabathil}},
  \bibinfo{author}{\bibfnamefont{E.C.}~\bibnamefont{Clark}},
  \bibinfo{author}{\bibfnamefont{A.}~\bibnamefont{Kress}},
  \bibinfo{author}{\bibfnamefont{D.}~\bibnamefont{Schuh}},
  \bibinfo{author}{\bibfnamefont{M.}~\bibnamefont{Bichler}},
  \bibinfo{author}{\bibfnamefont{G.}~\bibnamefont{Abstreiter}},
  \bibnamefont{and} \bibinfo{author}{\bibfnamefont{J.J.}~\bibnamefont{Finley}},
  \bibinfo{journal}{Phys. Rev. Lett.} \textbf{\bibinfo{volume}{94}},
  \bibinfo{pages}{057402} (\bibinfo{year}{2005}).

\bibitem[{\citenamefont{Williamson et~al.}(2000)\citenamefont{Williamson, Wang,
  and Zunger}}]{williamson00}
\bibinfo{author}{\bibfnamefont{A.~J.} \bibnamefont{Williamson}},
  \bibinfo{author}{\bibfnamefont{L.-W.} \bibnamefont{Wang}}, \bibnamefont{and}
  \bibinfo{author}{\bibfnamefont{A.}~\bibnamefont{Zunger}},
  \bibinfo{journal}{Phys.\ Rev.\ B} \textbf{\bibinfo{volume}{62}},
  \bibinfo{pages}{12963} (\bibinfo{year}{2000}).

\bibitem[{\citenamefont{Keating}(1966)}]{keating66}
\bibinfo{author}{\bibfnamefont{P.~N.} \bibnamefont{Keating}},
  \bibinfo{journal}{Phys. Rev} \textbf{\bibinfo{volume}{145}},
  \bibinfo{pages}{637} (\bibinfo{year}{1966}).

\bibitem[{\citenamefont{Wang and Zunger}(1999)}]{wang99b}
\bibinfo{author}{\bibfnamefont{L.-W.} \bibnamefont{Wang}} \bibnamefont{and}
  \bibinfo{author}{\bibfnamefont{A.}~\bibnamefont{Zunger}},
  \bibinfo{journal}{Phys.\ Rev.\ B} \textbf{\bibinfo{volume}{59}},
  \bibinfo{pages}{15806} (\bibinfo{year}{1999}).

\bibitem[{\citenamefont{Szabo and Ostlund}(1989)}]{Szabo89}
\bibinfo{author}{\bibfnamefont{A.}~\bibnamefont{Szabo}} \bibnamefont{and}
  \bibinfo{author}{\bibfnamefont{N.~S.} \bibnamefont{Ostlund}},
  \emph{\bibinfo{title}{Modern Quantum Chemistry}}
  (\bibinfo{publisher}{McGraw-Hill}, \bibinfo{address}{New York},
  \bibinfo{year}{1989}).

\bibitem[{\citenamefont{Franceschetti et~al.}(1999)\citenamefont{Franceschetti,
  Fu, Wang, and Zunger}}]{franceschetti99}
\bibinfo{author}{\bibfnamefont{A.}~\bibnamefont{Franceschetti}},
  \bibinfo{author}{\bibfnamefont{H.}~\bibnamefont{Fu}},
  \bibinfo{author}{\bibfnamefont{L.-W.} \bibnamefont{Wang}}, \bibnamefont{and}
  \bibinfo{author}{\bibfnamefont{A.}~\bibnamefont{Zunger}},
  \bibinfo{journal}{Phys.\ Rev.\ B} \textbf{\bibinfo{volume}{60}},
  \bibinfo{pages}{1819} (\bibinfo{year}{1999}).

\bibitem[{\citenamefont{Resta}(1977)}]{resta77}
\bibinfo{author}{\bibfnamefont{R.}~\bibnamefont{Resta}},
  \bibinfo{journal}{Phys.\ Rev.\ B} \textbf{\bibinfo{volume}{16}},
  \bibinfo{pages}{2717} (\bibinfo{year}{1977}).

\bibitem[{\citenamefont{Bennett et~al.}(1996)\citenamefont{Bennett, Bernstein,
  Popescu, and Shumacher}}]{bennett96}
\bibinfo{author}{\bibfnamefont{C.}~\bibnamefont{Bennett}},
  \bibinfo{author}{\bibfnamefont{H.}~\bibnamefont{Bernstein}},
  \bibinfo{author}{\bibfnamefont{S.}~\bibnamefont{Popescu}}, \bibnamefont{and}
  \bibinfo{author}{\bibfnamefont{B.}~\bibnamefont{Shumacher}},
  \bibinfo{journal}{Phys. Rev. A} \textbf{\bibinfo{volume}{53}},
  \bibinfo{pages}{2046} (\bibinfo{year}{1996}).

\bibitem[{\citenamefont{Bester et~al.}(2005)\citenamefont{Bester, Zunger, and
  Shumway}}]{bester05b}
\bibinfo{author}{\bibfnamefont{G.}~\bibnamefont{Bester}},
  \bibinfo{author}{\bibfnamefont{A.}~\bibnamefont{Zunger}}, \bibnamefont{and}
  \bibinfo{author}{\bibfnamefont{J.}~\bibnamefont{Shumway}},
  \bibinfo{journal}{Phys. Rev. B} \textbf{\bibinfo{volume}{71}},
  \bibinfo{pages}{075325} (\bibinfo{year}{2005}).

\bibitem[{\citenamefont{He et~al.}(2005)\citenamefont{He, Bester, and
  Zunger}}]{he05b}
\bibinfo{author}{\bibfnamefont{L.}~\bibnamefont{He}},
  \bibinfo{author}{\bibfnamefont{G.}~\bibnamefont{Bester}}, \bibnamefont{and}
  \bibinfo{author}{\bibfnamefont{A.}~\bibnamefont{Zunger}},
  \bibinfo{journal}{cond-mat/} \textbf{\bibinfo{volume}{0503492}},
  \bibinfo{pages}{} (\bibinfo{year}{2005}).

\bibitem[{\citenamefont{Edmiston and Ruedenberg}(1963)}]{edmiston63}
\bibinfo{author}{\bibfnamefont{C.}~\bibnamefont{Edmiston}} \bibnamefont{and}
  \bibinfo{author}{\bibfnamefont{K.}~\bibnamefont{Ruedenberg}},
  \bibinfo{journal}{Rev. \ Mod. \ Phys.} \textbf{\bibinfo{volume}{35}},
  \bibinfo{pages}{457} (\bibinfo{year}{1963}).

\bibitem[{\citenamefont{Craig and Walmsley}(1968)}]{Craig68}
\bibinfo{author}{\bibfnamefont{D.~P.} \bibnamefont{Craig}} \bibnamefont{and}
  \bibinfo{author}{\bibfnamefont{S.~H.} \bibnamefont{Walmsley}},
  \emph{\bibinfo{title}{Excitons in molecular crystals}}
  (\bibinfo{publisher}{W. A. Benjamin, INC.}, \bibinfo{address}{New York},
  \bibinfo{year}{1968}).

\end{thebibliography}
\end{document}